\documentstyle[12pt]{article}
\begin{document}
\begin{titlepage}
\author{Brukhanov ~A.A., Skulachev~D.P., Strukov~I.A., Konkina~T.V.}
\title{COBE Data Spatial--Frequency Analysis and \\
       CMB Anisotropy Spectrum.\footnote{accepted to Pis'ma v
Astronomicheskij Zhurnal, 1996, v.22, in Russian}}

\maketitle

\begin{center}
\it Space Research Institute, Moscow, Russia \newline
E-mail: istrukov@esoc1.iki.rssi.ru, virgo@relict.iki.rssi.ru$\;$
\end{center}
\hyphenation{ per-tur-ba-tion  qua-dru-po-le
con-cen-tra-ted tem-pe-ra-tu-res pa-ra-me-ters co-va-ria-tion
}

\begin{abstract}

We discuss the problem of CMB spectrum corruption during Galactic
emission removing. A new technique of spatial--frequency data
reduction is proposed. The technique gives us a possibility to
avoid a spatial harmonics nonorthogonality. The proposed
technique is applied to the two-year COBE DMR sky maps.

We exclude the harmonics with l=7, 9,13, 23 and 25 as having
anomalous statistics noise behavior. One shows that
procedure do not give systematic errors, if the data are
statistically regular.

The spectral parameter of the power spectrum of primordial
perturbation $n=1.84 \pm 0.29$ and quadrupole moment $Q_2=15.22
\pm 3.0$ are estimated. The power spectrum estimation results are
inconsistent with the Harrison-Zel`dovich $n=1$ model with the
confidence 99\%. It is shown a necessity of an increasing a
survey sensitivity to reach a more reliable estimation of the
cosmological signal.

\end{abstract}

\end{titlepage}

\section{Introduction.}

A large scale cosmic microwave background (CMB) anisotropy carries
the information about a primordial metric perturbation amplitude
and spectrum, it shape reflects the Universe evolution and
structure. To encrypt the information one analyzes a CMB
anisotropy spatial spectrum and compares it with a theoretical
models with different types of the Universe evolution scenario.
A sensitivity rising in the CMB investigations caused a detection
of the anisotropy. As a result one may find prerequisites of a
more detail examination of the physical conditions involved in the
Universe birth and evolve. Being high sensitive and multi-frequent
COBE experiment (Smoot et al., 1992, Bennett et al., 1994) gives a
possibility to careful expanding of our knowledge about the CMB
structure and spectrum.

Unfortunately, the Galaxy radiation predominates even at a
millimeter wavelength range. This rises a problem of the
radiation filtering. The most Galaxy radiation is coming from
rather narrow region located along the Galactic plane. The main
method to exclude the radiation is to "cut out" the Galactic plane
vicinity and to analyze the rest regions on the sky map (~Strukov
et al.,~1991a, 1991b, Smoot et al.,~1992, Bennett et al.,~1994,
Gorski,~1994, Gorski et al.,~1994, Wright et al.,~1994b). This
method is well known but it works perfectly only if variance data
analysis is applied. If we use the Galactic "cut out" in the case
of spatial spectrum analysis we lose a spherical functions
orthogonality on the rest sphere part. As a result the harmonics
begin to influence each to other. The problem may be solved partly
using a new basis which is orthogonal over the rest part of the
sphere (~Gorski,~1994) or is orthogonal only to a monopole and
dipole because the excluding of these two components effects the
most spectrum modification. The spectra obtained this ways may
differ from a real one (Bunn et al.,~1994) and may cause the
essential different conclusions when interpreting the results.

A modern multi-frequency experiments seemingly allow us to
separate the blackbody CMB radiation component from the frequency
dependent Galaxy foreground. But variations of the radiation
spectral composition prevent the procedure from being performed
with a desired accuracy.

We propose the method of the spatial spectrum analysis which
allows to reach the most spatial resolution without any loss in
the reciprocal harmonics orthogonality. The method implies that
the CMB filtering stage is conducted not before but after the
radio map is expanded to spherical harmonics.

There are two filtering stages in the proposed process. The first
stage is a spatial filtering one when some components related with
the Galactic plane and the Galaxy center are excluded from the
total spatial spectrum. By this expedient the Galactic radiation
magnitude is essentially reduced. And only after that, during the
second filtering stage a "frequency cleaning" procedure is used.
At this stage the CMB anisotropy is separated from the Galactic
radiation using several frequency channels and some a priori
Galaxy radiation model. The both stages nowhere are associated
with any signal harmonics orthogonality losses.

The main method difficulty may be found at the second stage and is
bound up with the problem of a correct Galactic radiation model
selection. But using the first stage filtering we strongly reduce
the effect of the model uncertainty on the net result.

As an initial data we used public released COBE DMR 2-years all sky
maps from 31.5, 53, and 90 GHz frequency channels. In addition we
exploited a 19.2 GHz balloon data kindly provided by the authors
of the survey.

\section{ The Method.}

A data reduction process practically always is accomplished by
some types of the data censoring like weighing, excluding outliers,
corrections for systematic errors, trends and so on. A censor
method selection is based on a priori information about the
analyzing sample. The censoring is aimed at the improvement of the
estimation stability and at the approaching the real data statistic
parameters to the used a priori models.

In our case the values on the observed radio map is a sum
of a cosmological signal we try to measure, a receiver noise, the
Galactic radiation, and perhaps some signal related with neglected
systematic effects. It is the cosmological signal that we are
interested in, so the other components remain to be carefully
estimated, significantly reduced or to be completely excluded.
In an ideal case (no Galactic radiation and no systematic effects)
the procedure would not cause a significant change in the
cosmological signal parameters.

In a Galactic coordinate system one may find a Galactic emission
concentrated near the Galactic plane and around the Galaxy center.
The radiation may be described using only a few spatial harmonics.
If these harmonics are excluded the Galactic radiation is
substantially reduced. In our case we excluded 23 harmonics only.
At the same time the rest spatial spectrum consist of 649
harmonics. If a signal is normal distributed the 649 harmonics are
quite sufficient for a stable signal spectrum parameter estimation.

To estimate the signal parameter adequately it is absolutely
necessary to know (or to estimate) exactly a noise which may be
found in any analyzed data. The most problem in the procedure is
the presence of anomalous magnitudes in the noise spectrum. If the
signal/noise ratio is not good enough the anomalies may cause a
significant variation in the parameters to be estimated.

We emphasize that a cause of the anomalies may be some residual
systematic errors. In this case it is very difficult to build up a
statistical model of the anomaly and the accuracy of the signal
estimation may drops below a point that can be tolerated.

If the number of the anomalous harmonics is not large in
comparison with the total signal harmonics, the radical solution of
the problem is an exclusion of the anomalous harmonics from the
analysis. If data are regular, the excluding some anomalous (in the
noise spectrum) harmonics is rather safe procedure.
That is because the cosmological signal and the noise are Gauss
type, and the signal parameters are weakly depended on both the
fact of excluding and the criteria of anomalous outliers selection
(in our case we rated to anomalous the outliers beyond the
3--sigma).

But if the neglected systematic errors are present in the data,
this procedure causes the significant improvement of the sample
regularity and causes the increasing of an estimation
efficiency. At the same time if the excluding process significantly
changes the estimated parameters, it may be a strong reason to
assume the data is not regular, and consequently may indicates
that it is necessary to use some type of procedure to reduce the
irregularity.

It should be noted that using data reduction methods which break a
spatial harmonics orthogonality causes a certain spectral
smoothing and may reduce or even may camouflage the abnormal
amplitudes. In this situation the risk of missing a really
existing anomaly is run, and the signal estimation process may be
accomplished with some additional errors.

The data reduction procedure we propose is the following. We use
COBE A and B channels (Wright et al.,~1994a) and produce the sum
and the difference radio-maps as a half-sum and a half-difference
of the A and B maps respectively. All data on the sum map are a
sum of a signal and a noise. All data on the difference map are
only the noise, the signal amplitude is zero.

We exclude from the sum map the monopole and dipole components,
and the spectral components cased by the Galaxy emission.
Then the sum map is additionally "cleaned" from the Galactic
radiation. For this purpose we use 19 GHz and 31 GHz maps to
determine the Galactic radiation spectral index. Then the 31 GHz
map is subtracted from 53 GHz map, but with a weight corresponding
the determined spectral index.

We assume the hypothesis the cosmological signal and the noise are
normal distributed. The noise spectrum shape is assumed to be
completely determined by the number of measurements in every point
on the sky. Based on this we calculate the noise parameters and
discover anomalous noise components. Then we use a Monte Carlo
simulation to estimate signal magnitude, signal spectrum
parameters, and the estimates confidence levels.
The latter procedure is applied both to a total analyzed spectrum
and to a censored spectrum with anomalous components excluded.
In addition we test the determined estimations for stability to
the procedure of quadrupole component subtracting.

\subsection{ Signal Analysis on the Sphere.}

The signal on the sphere may be represented as a spherical
harmonics expansion $Y_{l,m}(\theta, \varphi)$:
$$
\Delta T(\theta, \varphi)  = \sum a^m_l Y_{l,m}(\theta, \varphi)
$$
In this case the signal variance on the sphere $\sigma^2$ is
represented as:

$$
\sigma^2 = \frac{1}{4 \pi} \sum_l \sum_m (a_l^m)^2 = \sum_l \Delta
T_l^2 \,
$$
where $\Delta T_l^2$ is a $l$--th spherical harmonic power.

In any real experiment the analyzed signal $a^m_{l,signal}$ is
affected by a transfer function $W_l$ (thereafter we use
the COBE antenna transfer function (Wright et al.,~1994a) as
$W_l$) and some noise $a^m_{l,noise}$ is added:

$$
\Delta T (\theta, \varphi) = \sum (W_l a^m_{l,signal}+a^m_{l,noise})
Y_{l,m}(\theta, \varphi) \,
$$

The cosmological signal on the sphere is a sample of a random
process. In a standard cosmological scenario the inflation creates
adiabatic scalar and tensor Gauss fields. So the spectrum is
assumed to be a sample of a normal random process with
independent harmonics, $\langle a_l^m a_k^n \rangle
\sim \delta
(l,k) \delta (m,n)$.

If initial density perturbation spectrum
is $(\delta \rho / \rho)^2 = A k^n$,
one may find for the mean lowest multipoles in expansion
$\Delta T(\theta, \varphi) $ (Bond, Efstathiou, 1987):

\begin{equation}
\langle (a_l^m)^2 \rangle \sim \frac{\Gamma(l+(n-1)/2)
\Gamma((9-n)/2)}
			{\Gamma(l+(5-n)/2) \Gamma((3+n)/2)}
\label{approx_DTT}
\end{equation}

A consideration of other physical phenomena (~evolution effects,
acoustic oscillation, Silk effect etc.) causes a more complicated
spectrum (White et al.,~1994 and references therein). Hereafter we
approximate a real spectrum with the~(\ref{approx_DTT}). For a
quantitative spectrum description we use only two parameters:
a power spectrum index $n$ and quadrupole magnitude
$Q^2_{rms}=\langle \Delta T^2_2 \rangle$ as an amplitude.

A correlation function on the sphere
$C(\beta)=\langle(\Delta T(\vec q_1)) (\Delta
T(\vec q_2))\rangle $, $\vec q_1 \vec q_2=\cos(\beta)$
may be represented in terms of spherical
functions:  $$ C(\beta)=\frac{1}{4 \pi} \sum P_l(\cos \beta ) \sum_m
(a_l^m)^2 $$

Using the spherical harmonic expansion we may calculate the
correlation between two maps (for example between 31.5~GHz and
53~GHz maps):
$$
\rho({\rm MAP1, MAP2}) = \frac{ \sum_{l,m} a_l^m(1) a_l^m(2) }
      {\sqrt{ \sum_{l,m} (a_l^m(1))^2 \sum_{l,m} (a_l^m(2))^2 }}
$$

\subsection{ Spectrum Parameters Analysis.}

A cosmological model scenario gives us a prediction of a mean CMB
spectrum fluctuation. Experimenters deal with the single sample of
the scenario. In addition a noise is always present in a real
data. In this connection there is the problem of a signal
parameters estimation. The problem was discussed by Bunn et al.,
1994 and Sazhin et al.,~1995. Here we will not go into the
problem, but will touch upon the estimation not only the signal,
but the noise parameters too.

The maximum likelihood estimation is known to be the most
effective. It may be illustrated if the signal spectrum has a given
a priori shape.

Let the spectrum shape is $\langle (a_l^m)^2 \rangle =Q^2 F_{lm}$. If
the amplitude is $Q$, the likelihood function may be described as:
$$
f(a_{lm},Q|F_l) = \prod_{lm} \frac{1}{\sqrt{2 \pi Q^2 F_{lm}}}
 \exp \left( - \frac{ a_{lm}^2 } { 2 Q^2 F_{lm} } \right),
$$

The maximum likelihood estimation $Q_{\rm ML}$ may be found if
solve the equation maximizing the likelihood function:
$$
\frac{\partial \ln f(Q|F_l)}{\partial Q^2} = 0
$$
$$
Q^2_{\rm ML} = \frac {\sum (a^2_{lm}/F_{lm})}{ M }, \qquad
 M=\sum_{l,m} 1 \,
$$

where $M$ is the number of the analyzed harmonics.

The parameter is evident having $\chi^2_M$ distribution with $M$
degree of freedom, $Q^2_{\rm ML} \sim \chi^2_M$. The amplitude
estimation calculated using the variance on the sphere $ \sigma^2
$ is:
$$
Q^2_{\rm POWER}=\frac{\sum \Delta T^2_l}{\sum \langle \Delta
T^2_l \rangle |{Q=1}}
= \frac{\sigma^2}{\sigma^2_{th}|Q=1} \sim \chi^2_{N_{eff}} \, .
$$

It is clear that:
 $N_{eff}= \frac{(\sum_{lm} \Delta T^2_l)^2}{\sum_{lm} (\Delta
T^2_l)^2} \le  M$. So for the Harrison-Zel'dovich spectrum
$N_{eff} \approx 100$, $ M \approx 700$ пpи $l_{max} =25$.
The maximum likelihood estimation accuracy is about 5\,\% and the
total power estimation is about 14\,\%.

Using the approach we may determine the signal spatial spectrum
parameters in the presence of noise. Maximum likelihood
estimation offers to find a minimum of the functional:

\begin{equation}
\sum_{l,m} \ln(W_l^2 \langle(a_l(Q,n))^2 \rangle + \langle (a_{l,noise}^m)^2
\rangle ) + \frac{(b_l^m)^2}{W_l^2 \langle(a_l(Q,n))^2 \rangle +
\langle (a_{l,noise}^m)^2 \rangle}
\label{likhood}
\end{equation}

relative $Q$ and $n$ with given spectrum shape and measured signal
and noise sample spectra. Here $\langle(a_l(Q,n))^2 \rangle$ is
the model signal spectrum, depending upon desired parameters and
described as (\ref{approx_DTT}), $(b_l^m)^2$ is the measured spectrum
(we describe it further), and  $\langle(a_{l,noise}^m)^2 \rangle$
is a model noise spectrum~(\ref{sp_noise}), its amplitude is found
using the same approach but from the difference map .

In addition we are to take into account a procedure of the
Galactic radiation "cleaning" and a procedure of a conversion the
antenna temperatures into a thermodynamic scale. In all cases the
summation is taken over multipoles chosen for the analysis. The
necessary conditions for the described procedure is a knowledge
of the measured signal and noise spectra.

\subsection{ COBE Data Noise Analysis.}

There are a lot of papers concerning the COBE data noise analysis.
Bennett et al.,~1994, Lineweaver et al.,~1994 discussed the
correlation noise properties and showed a weak (~0.45\%~)
correlation at 60$^\circ$. So in practice we may assume the noise
is uncorrelated. The noise power estimation may be done using the
difference maps.

As a first approximation assume that the noise is white. However
the real observations have different accuracy on the sphere. The
COBE orbit configuration was so that the points near the ecliptic
poles were observed much longer than others. Let us consider the
methods to take into account the measurement accuracy is
not equal for different points.

Let the noise in every point on the sphere be normal with zero
mean and the variance
$\sigma^2(\theta,\varphi)=\sigma_0^2/N(\theta,\varphi)$,
where
$N(\theta,\varphi)$ is a number of measurements in the point with
the $\theta$, $\varphi$ coordinates. Then we may propose the
following noise spectrum model on the sphere. If there is a white
noise sample
$G=\sum n_l^m Y_l^m$, $\langle n_l^m
n_{l^
{\prime}}^{m^{\prime}} \rangle = 0$, $\langle (n_l^m)^2
\rangle =n^2=const$,
the true noise on the sphere is a product of $G(\theta,\varphi)$
(here, and only here $n$ is a magnitude of the noise power spectrum,
not a spectrum index),
and
$1/\sqrt{N(\theta,\varphi)}$:
$$2
 R(\theta,\varphi) = 1/\sqrt{N(\theta,\varphi)} G(\theta,\varphi) =
w(\theta,\varphi)G(\theta,\varphi)
$$

According to Peebls~(1980), assume $w(\Omega)Y_l^m(\Omega) =
\sum_{l^{ \prime} m^{\prime}} w_{l l^{\prime}}^{m m^{\prime}}
 Y_{l^{\prime}}^{m^{\prime}}(\Omega)$, where
$$
w_{l l^{\prime}}^{m m^{\prime}} = \int d \Omega w(\Omega)
 Y_{l^{\prime}}^{m^{\prime}}(\Omega) Y_l^m(\Omega) \
$$

For the true noise spherical harmonics expansion
$R=\sum a_{l,noise}^m Y_l^m$, coefficients are
 $a_{l,noise}^m=\sum_{l^{\prime} m^{\prime}} w_{l l^{\prime}}^{m
m^{\prime}}
 n_{l^{\prime}}^{m^{\prime}}$.

Taking into account noncorrelatedness of white noise harmonics
a covariation of $a_{l,noise}^m$ may be written as:
$$
\langle a_{l,noise}^m a_{l^*,noise}^{m^*} \rangle =\sum_{l^{\prime}
m^{\prime}}
 \sum_{l^{\prime\prime} m^{\prime\prime}}
 w_{l l^{\prime}}^{m m^{\prime}}
 w_{l^* l^{\prime \prime}}^{m^* m^{\prime\prime}}
 \langle n_{l^{\prime}}^{m^{\prime}} n_{l^{\prime \prime}}^{m^{\prime
\prime}} \rangle = \\ n^2 \sum_{l^{\prime} m^{\prime}} w_{l
l^{\prime}}^{m m^{ \prime}} w_{l^* l^{\prime}}^{m^* m^{\prime}}
$$

And finally noise power spectrum is:
\begin{equation}
\langle|a_{l,noise}^m|^2 \rangle =n^2 \sum_{l^{\prime} m^{\prime}}
|w_{l l^{\prime}}^{m
m^ {\prime}}|^2 = n^2 \int d \Omega  w^2(\Omega)|Y_l^m(\Omega)|^2
\label{sp_noise}
\end{equation}

We use not only unit or zero weights, but unrestricted values.
Because of that our equation differs from Peebls's one in the
squared weights.

Then, using maximum likelihood approach to the difference maps we
may calculate the amplitude of the spectrum given above.

The power spectrum model make it possible to analyze the noise
spectrum measured from the difference maps. The analysis shows that
in general the noise spectrum is well described both the model
(\ref{sp_noise})
and the white noise model. It is due to a rather perfect COBE
coverage of the celestial sphere. But the same analysis shows for
the both models the existence of harmonics with an excess power.
So in the noise spectrum of 31~GHz map one may find for $l=13$
the power is greater than predicted from~(\ref{sp_noise}). The
power excess is 3.4$\sigma_{l=13}$, where predicted multipole
power variation $\sigma_l$ is calculated basing on the model and
on the degrees of freedom for the analyzed component. Further we
use a censoring of the data to exclude this and similar components
from the analysis.

The noise spectral component with $l=25$ has anomalous high correlation
between 31 and 53 GHz maps. After subtracting the Galaxy
radiation the excess power is 3.3$\sigma_{l=25}$. Components with
$l=25$ were excluded from the analysis too.

We find a significant power excess of several noise components on
53 GHz map. We exclude the components with power excess more than
3$\sigma_{l}$. After that we determine the spectrum parameters for
the rest components and the procedure is repeated until the
spectrum is free from the abnormal components. Thus we excluded
the components with $l=7,\, 23,\, 9$ having the initial power
excess of 3.11$\sigma_{l=7}$, 2.63$\sigma_{l=23}$, and
2.31$\sigma_{l=9}$.

In the final analysis we use spectral components from $l_{min}=2$
to $l_{max}=22$, except components $l=7,\, 9,\, 13$.

Such censoring effects only a weak power spectrum distortion if
the signal amplitude distribution is normal. The matter is discovered by
numerical modeling, the results are presented in Chapter 3.

During signal spectrum analysis we ignore a noise spectrum
components correlation and use the spectrum shape only. Further
modeling shows the same results if we use a pure white noise.

\subsection{Selection of the Galactic Radiation Model}

Our approach is based on the spatial data filtering which reduces the
influence of the Galactic spectrum index uncertainty. The accepted
Galactic radiation model for the analyzed frequency region is
assumed to involve a synchrotron component with frequency
dependence as
$T=T_0\nu^{\alpha}$, $\alpha=-3\pm 0.2$
and a bremsstrahlung component with spectral index $\alpha=-2.1$
(Bennett et al.,~1992). If there are the space regions of a
different nature of radiation on the line of sight, it may cause a
spatial variation of the spectral index $\alpha$.
Let us try to estimate roughly the contribution of such variation
if we use some effective but constant value $\alpha_0$. Let the
true value of the spectral index be $\alpha=\alpha_0+\Delta
\alpha$. After the frequency depended part is excluded from the
data $T_1$ и $T_2$ of two frequency channels, one may derive:
$$
 \delta T = T_1 \left(\frac{\nu_2}{\nu_1}\right)^{\alpha_0+\Delta
\alpha} - T_1
\left(\frac{\nu_2}{\nu_1}\right)^{\alpha_0} \approx
T_2 \left[ \left(\frac{\nu_2}{\nu_1}\right)^{\Delta \alpha}
-1 \right] \approx
T_2 \Delta \alpha \ln \left(\frac{\nu_2}{\nu_1}\right)
$$

The residual signal power caused by the spectral index variation
$\sigma^2_{\rm VAR}= \Delta \alpha^2 (\ln\nu_2 /\nu_1)^2
\sigma^2_{\rm MAP2} $. For the frequencies 31.5 and 53 GHz the
residual RMS is smaller than
\linebreak[4] $0.52 \Delta \alpha \sigma_{\rm MAP2}$. It can be
seen that the less is Galactic radiation contribution, the less is
the residual RMS on the more high frequency radio map.

It is evident the most Galactic radiation is concentrated near the
Galaxy plane and near its center. If the spatial spectrum is
represented in Galactic coordinate system the most power may be
found in the components with $m=0$ for the even multipoles (Galaxy
plane) and with $m=1$ for the odd multipoles (Galaxy center). So
if we exclude this components the influence of the spectral index
variation may be reduced drastically. At the same time this
procedure does not break the multipoles orthogonality.

Table~\ref{tab_gal} shows some results obtained after the spectrum
analysis from $l_{min} =2$ to $l_{max}=25$.

In addition one may find at the last row the difference between
53~GHz map and the 31~GHz map scaled with a spectral index $\alpha
= -2.15$ after spatial filtering. Data are represented in antenna
temperatures. One can see the resultant $\sigma_{Sky}$ may be
explained either by a cosmological signal existence or by the
spectral index variation within $\Delta \alpha \approx 0.5$.

Unfortunately the available data give no way to estimate the
variation with required precision. For more reliable conclusions it
is necessary to have the sensitivity many times better than COBE
instrument has.

We tried to estimate the spectral index variation crude. For this
purpose we use both the COBE data and 19.2 GHz survey (Boughn et
al.,~1992). We has converted latter data to a COBE beam shape
(Wright et al.,~1994a). After that we analyzed the frequency
dependence of the most intensive spherical harmonics and
determined the accuracy of the estimation.

The data are shown in Table~\ref{indx} where one can see
correspondingly: analyzed frequency ranges, spectral index
$\alpha$ for desired signal/noise ratio equal to 5, spectral index
variation $\Delta\alpha_5$, and the predicted spectral index
variance $\Delta\alpha_N$, calculated basing on the noise
analysis. In addition there are shown the spectral index $\alpha_W$
and its variation weighted-mean for all harmonics with
$m\ne0$ for $l=2k$ and with $m\ne1$ for $l=2k+1$.

It may be noticed the absolute value of the spectral index
$\alpha_{19-31}$ is regular lower than $\alpha_{31-53}$. The
contribution to the total spectrum of the spectral components with
signal/noise ratio equal to 5 is more than 40/
found the only spectral component with the spectral index like
a synchrotron Galactic radiation has. This is $a_{2,2}$ component
with the spectral index
$\alpha_{19-31}=-2.86\pm 0.15$, $\alpha_{31-53}=-2.92\pm 0.22$.

The different maps correlation analysis gives the mean spectral
index estimate in the region $\alpha=-(2.2 \div 2.3)$.

So we may assume the hypothesis the spectral index is constant and
its measured variations are caused only by the instrumental noise.

On the basis of the analysis we can conclude for the most
significant Galactic spectral components (with the exception of
$a_{2,2}$~) the Galactic radiation model may be assumed to be
one-component, with the radiation cased only by an ionized
hydrogen. So the $\Delta\alpha$ variations may be taken as zero.

Now we may determine the measured spectrum

$$
b_l^m = \left(
 k(53) - k(31) \left( \frac{53}{31.5}\right)^{-2.15}
						 \right)^{-1}
\left(
a_l^m(53\,GHz) - a_l^m(31.5\,GHz)
	  \left( \frac{53}{31.5} \right)^{-2.15}
 \right) \, ,
$$

where  $k(31)=1./1.025724$,  $k(53)=1./1.074197$ are the
scale coefficients to transform the thermodynamic temperatures to
antenna temperatures for 31.5~GHz and 53~GHz, $a_l^m(31.5\,GHz)$,
$a_l^m(53\,GHz)$ are multipole coefficients for 31.5~GHz and 53~GHz
maps.

Unfortunately, the COBE sensitivity is not good enough and the
accuracy of the estimation is determined by the instrumental
noise. We are pinning our hopes on the "Relict-2" space
experiment. "Relict-2" instrument is preparing now, it has an
order better sensitivity than COBE has, and it will give the
possibility to analyze the Galactic spectral index radiation with
high reliability and so to separate the cosmological signal from
the Galactic radiation with high accuracy.

\section{Results}

As a result of the analysis we obtained the spatial spectrum
parameters estimates under the assumption that Galactic radiation
is one-component. For the quadrupole we found:
$$
Q_2=15.22 \pm 3.0  \mu{\rm K}
$$
and for the power spectrum index:
$$
n=1.84 \pm 0.29
$$
We analyzed spectral components from $l_{min}=2$ to $l_{max}=22$
if $m\ne 0$ for $l=2k$, and $m\ne 1$ for $l=2k+1$,
and (due to the noise anomalies) if $l \ne 7,9,13$.

The accuracy of the obtained parameters is derived using
Monte-Carlo simulation. For all estimates we show RMS errors.
The procedures of difference and sum maps producing and of noise
parameters determining are included in the Monte-Carlo simulation.
The simulation also takes into account the noise, its variations, and
a cosmic variance.

We find the total power measured in chosen spatial frequency
window as $(70.28\,\mu{\rm K})^2$, noise estimate as
$(56.76\,\mu{\rm K})^2$, and on-sphere signal estimate as
$(41.44\,\mu{\rm K})^2\pm(14.68\,\mu{\rm K})^2$. Total number of
analyzed harmonics is $M=446$.

Table~\ref{tbl_n} shows the results. In addition we include into
the Table~\ref{tbl_n} the data obtained with anomalous harmonics,
and with excluded quadrupole.

We tested an effect of the harmonics excluding procedure on the
signal estimation. In the case of the noise outliers caused by
systematic errors and if all harmonics are orthogonal the
excluding does not effect the estimation bias, but may only
increase the errors in comparison to non-censored data. Simulation
shows the $Q_2$ and $n$ variations are 0.9 and 0.11 respectively.

On the other hand, if the outliers are pure stochastic (i.e. we
are dealing with a low-probability noise sample) the noise
estimation bias may arise and, as a result may arise the signal
estimation bias. We simulated this occasion for the signal and
noise with the given spectrum and to examine noise outliers
stronger than 2.3$\sigma$. After that we calculated the signal
estimates both for censored and non-censored spectra.

After 1228 simulations we have found 500 events with the outliers.
The corresponding estimates are: $\langle
n_{censor}-n_{full}\rangle=0.07\pm 0.08$ $\langle
Q_{censor}-Q_{full} \rangle=-0.4\pm 0.83$. It should be noted
that the obtained estimates really are upper bounds rather than
two-sided limits. It is because we exclude the components, which
then are used in some linear combination and thus are additionally
normalized. It may, in general, reduce the effect of spectrum
censoring.

COBE data analysis shows (see Table~\ref{tbl_n}) that the
procedure of the anomalous harmonics excluding effects the
significant change in $n$. If the instrumental noise is completely
random the probability to obtain so large difference is smaller
than $0.02\%$. One may see that the applying the anomalous
harmonics censoring to COBE data causes a dramatically decreasing
of the strange dependence of the results on whether or not the
quadrupole is excluded. It may be an indirect support to the
necessity of proposed data censoring.

Basing on the obtained parameters we may predict the signal in
an experiment like it is conducted in Tenerife (Hancock et
al.,~1994). Assuming $5.5^{\circ}$ beam, $8.1^{\circ}$ antennae
separation, and 3-point method of observation i.e. $\Delta T = T_0
-0.5(T_1 + T_2)$, we may calculate the following data for the
spectrum~(\ref{approx_DTT}) type using previously obtained $Q_2$
and $n$, excluding components with anomalous noise and applying
the spatial filtering:
$\sigma_{Tenerif} = 54.82 \, \mu{\rm K}$.

The same analysis, but without excluding anomalous components
shows two estimates:

$\sigma_{Tenerif} = 37.22 \, \mu{\rm K}$ with quadrupole included,
and:
$\sigma_{Tenerif} = 34.47 \, \mu{\rm K}$ with quadrupole excluded.

We do not use a $4^{\circ}$ binning in the analysis. The binning
may decrease the obtained data in some degree. The binning is used
in Tenerife experiment and the results are (Hancock~S. et
al.,~1994):

$\sigma_s=49\pm 10 \, \mu{\rm K}$ for 33 GHz channel, and
$\sigma_s=42\pm 9 \, \mu{\rm K}$ for a sum of the 15 and 31 GHz
channels. So the cosmological signal spectrum with the parameters
we obtained is in a good agreement with the observations which are
more sensitive than COBE for high spatial harmonics.

\section{ Conclusions.}

The systematic effects exclusion problem is met practically in any
experiment. In the case of CMB anisotropy observation the Galaxy
radiation is the primary effect. The method usually used to
suppress the radiation is to "cut off" the Galaxy plane region. It
causes the problem of the monopole and dipole exclusion. The
latter in its turn effects the additional systematic errors but at
the less level.

The orthogonal basis when coupled with the spatial-frequency
filtering of the Galactic radiation allows to avoid the systematic
errors mentioned above. In addition the proposed approach make it
possible to analyze and to exclude more fine effects usually
caused errors in spectrum parameters measurements. If an
instrumental noise is normal the data censoring could not cause a
significant difference between the spectral parameters derived
from censored and non-censored data sets. Being detected the
difference may be result from either neglected residual effects or
a noise nonnormality. Being in the context of the normal
distribution function we are forced to exclude the spectral
components with an anomalous noise behavior.

The cosmological spectrum parameters we obtained differ in some
degree from the parameters derived in previous works used the same
initial COBE data. The difference is due to the anomalous noise
harmonics exclusion rather than spatial--frequency Galactic
radiation filtering. In its turn the harmonics exclusion is
possible because of the harmonics orthogonality conservation.

A number of investigators (Table~\ref{rez}) announced the strange
end result sensitivity to a quadrupole component. Most likely it
is also attributed to anomalous noise harmonics influence.

The results based on 2-year COBE data and obtained by several
authors are shown in Table~\ref{rez}. In addition we would like to
remind the result of $n=1.7$, derived by~(Hancock~S. et al., 1994)
after a comparison of 1-year COBE and Tenerife data.

So we obtained the estimate of $n$ more precise and somewhat
higher than it is derived by other investigators.

To the best of our knowledge, nobody has investigated the COBE
noise (A--B) maps in detail. Gorski et al.~(1994) used only the
mean noise parameters assumed they are normal. Wright et
al.~(1994b) used the noise spectrum, but the used basis was not
complete orthogonal. Bennett et al.~(1994) worked by the help of a
correlation function but only on the part of the sphere. In the
latter two cases the basis function orthogonality is lost and it
is impossible to analyze noise spectrum in detail.

The next result we obtained and that differs from previously
published is the rejection of the trivial (with $n=1$)
Harrison-Zel'dovich spectrum on the confidence level of~99\%.

As a result there are very unlikely the models with high
$\Lambda$--term and a more probable are the open models with
$\Omega <1$, if we assume the initial density perturbations are
determined by the power law index $n=1$~(Kamionkowski~M.,
Spergel~D.N.,~1994). Our result supports the existence of a
barionic entropy models. At the same time there are not ruled out
the models having $n>1$ in analyzed scales and having more
complicated inflation potential (Starobinsky~1992, White et
al.,~1994 and references therein).

It must be emphasized that by now the spectrum parameters
determining accuracy is not enough to draw more deep inferences.
Moreover we can see the obtained results being very sensitive to the
accuracy of the noise determining and to the used specific
estimation procedure.

We hope the 4-year COBE investigation circle will be accessible in
the near future. Unfortunately even the 4-year data will not
improve the situation drastically. First, it is necessary to
increase the instrumental sensitivity as a minimum an order. This
will be reached in the planned experiment "RELICT-2". So the
accuracy of the $n$ estimation will be about 5-7\% and will be
determined the cosmic variance rather than instrumental noise
(Sazhin et al.,~1995). Second, it is necessary to enhance the
anisotropy measurement angle resolution coupled with a sky
coverage increasing (Scott et al.,~1994). Unfortunately, the
modern middle and small scale investigations give us the
information only about a few point on the sky and so the results
are strongly contaminated by the sample variance.

Authors thank Boughn~S.P., Cheng~E.S., Cottingham~D.A., and
Fixen~D.J. for providing the 19~GHz data. The COBE data sets were
developed by the NASA/GSFC under guidance of COBE Science Working
Group and were provided by the NSSDC. This work was supported
partly by RFFI grants N93-02-930, 93-02-931, ISF grants N~MO6000,
MO6300. We would like to thank A.~Klypin and M.~Sazhin for helpful
discussions and useful comments.

\pagebreak[4]
\section{ References.}
\begin{list}{}{\listparindent=-1cm}

\item Bunn et al.~(Bunn~E., Hoffman~Y., Silk~J.)//
Astrophys.J.~1994.~V.425.~P.359.

\item Bennett et al.~(Bennet~C.L., Smooth~G.F., Hinshaw~G., Wright~E.L.
et.al.)//
Astrophys.J.(Letters) 1992.~V.396.~L7.

\item Bennett et al.~(Bennet~C.L., Kogut~A., Hinshaw~G., Banday~A.J. et.al.)//
 Astrophys.J. 1994.~V.436.~P.423.

\item Boughn et al.~(Boughn~S.P., Cheng~E.S., Cottingham~D.A.,
Fixen~D.J.)// Astrophys.J. (Letters) 1992.~V.391.~L49.

\item Bond, Efstathiou~(Bond~J.R., Efstathiou G.)//
 Monthly~Not.Roy.Astron.Soc. 1987.~V.226,~P.655

\item Gorski~(Gorski Krzysztof.~M.)//
 Astrophys.J.(Letters)~1994.~V.430.~L85.

\item Gorski et al.~(Gorski~K.M., Hinshaw~G., Banday~A.J., Bennet~C.L.
et.al.)//
 Astrophys.J.(Letters)~1994.~V.430.~L89.

\item Kamionkowski, Spergel~(Kamionkowski~M., Spergel~D.N.)//
 Astroph.J., 1994, V.432, P.7-16

\item Lineweaver~(Lineweaver~C.H., Smooth~G.F., Bennet~C.L.,
 Wright~E.L. et.al.)//
 Astrophys.J. 1994.~V.436.~P.452.

\item Peebls~(Peebls P.J.E.)//	The Large-Scale Structure of the Universe,
Princeton:~Princeton Univ.Press,~1980.

\item Wright et al.~(Smooth~G.F., Kogut~A., Hinshaw~G., Tenorio~L.
et.al.)// Astrophys.J.~1994а.~V.420.~P.1

\item Wright et al.~(Wright~E.L., Smooth~G.F., Bennet~C.L., Lubin~P.M.)//
 Astrophys.J. 1994б.~V.436.~P.443.

\item White et al.~(White~M., Scott~D., Silk~J.)//
Ann. Rev. Astron. and~Astrophys. 1994.~V.32.~P.319.

\item Hancock et al.~(Hancock~S., Davies~R.D., Lasenby~A.N., Gutierrez
de~la Cruz~C.M. et.al.)// Nature.~V.367.~P.333.

\item Sazhin et al.~ (Sazhin~M.V., Brukhanov~A.A, Strukov~I.A.,
Skulachev~D.P.) // Pis'ma v astronomichesky zhurnal, 1995, v.21,
p.403 (in Russian, in English see Astronomy Letters ( Soviet Astronomy
Letters), 1995, v.21, p.358-365)

\item Scott et al.~(Scott~D., Srednicki~M., White~M.)//
Astrophys.J.(Letters)~1994. V.421.~L5.

\item Smoot et al.~(Smoot~G.F., Bennett~C.L., Kogut~A.,
Wright~E.L., et.al.)// Astrophys.J.(Letters)~1992.~V.396.~L1.

\item Starobinsky (Starobinsky~A.A)~// Pis'ma v zhurnal experim. i
teoret. fiziki, ~1992.~v.55.~p.477 (in Russian)

\item Strukov et al.~(Strukov~I.A., Brukhanov~A.A, Skulachev~D.P.,
Sazhin~M.V.)// Pis'ma v astronomichesky zhurnal~1992a.~v.18.~p.387
(in Russian, in English see Soviet Astronomy Letters,
 1992, v.~18, p.~153~)

\item Strukov et al.~(Strukov~I.A., Brukhanov~A.A, Skulachev~D.P.,
Sazhin~M.V.)// Monthly~Not. Roy. Astron. Soc. 1992b.~V.258,~P.37p.

\item Sugiyama, Silk~(Sugiyama~N., Silk~J.)//
Phys.Rev.Letters.~1994.~V.73.~P.509.

\end{list}
\pagebreak[4]
\section{Table Caption}
\begin{list}{}{\listparindent=-1cm}

\item {\bf Table~\ref{indx}.} Mean spectral indexes and its
variations determined for spherical harmonics at different frequency
ranges.
The first column -- the frequency ranges where the spectral index
$\alpha$ is analyzed.
The second column -- spectral index determined from the most
significant spectral components with the signal/noise ratio lager
than 5, $\alpha_5$.
The third column -- its variation, $\Delta\alpha_5$.
The forth column -- spectral index variance predicted from noise
analysis, $\Delta\alpha_N$.
The fifth column -- the mean weighted spectral index $\alpha_W$
for all harmonics with $m\ne0$ for $l=2k$, and $m\ne1$ for
$l=2k+1$.
The sixth column -- its variance~$\Delta\alpha_W$.

\item {\bf Table~\ref{tab_gal}.} Comparative result of an
influence of spatial filtering (i.e. excluding the components with
$m=0$ for $l=2k$ and $m=1$ for $l=2k$ from $l=2$ to $l=25$
inclusively) to a signal amplitude for different frequencies.
The last row shows the signal amplitude after the spatial --
frequency filtering assuming the Galactic radiation spectral index
as corresponded to ionized hydrogen. $\sigma_{(A+B)/2}$
-- amplitude, calculated as a half-sum of two maps (i.e. signal plus
noise), $\sigma_{(A-B)/2}$ -- noise, calculated as a
half-difference of two maps, $\sigma_{Sky}$ -- signal
amplitude estimation on the sky.

\item {\bf Table~\ref{tbl_n}.} Spectrum analysis after
spatial--frequency Galactic radiation filtering. The influence of
quadrupole excluding is shown. In addition it is shown the results
after noise anomalous harmonics are excluded. The first row is
corresponded to spectrum from $l=2$ to $l=22$ with the abnormal
harmonics excluded. The second row -- the same but in addition the
quadrupole is excluded. The third row shows the results if all
harmonics from $l=2$ to $l=25$ are used. The forth row shows in this case
the influence of quadrupole exclusion.
$\sigma_{A+B}^2$ -- total measured power within given spatial
window, $\sigma_{A-B}^2$ -- noise power estimation obtained
within the same window from a difference map,
$\sigma_{Sky}^2$ -- cosmological signal power estimation,
($ \sigma_{Sky}^2=\sigma_{A+B}^2 - \sigma_{A-B}^2 $ ), $Q_2$, $n$
-- estimations of parameters for spectrum~(\ref{approx_DTT}), $M$
-- total number of analyzed harmonics. In all cases we assume that
$m\ne0$ for $l=2k$ and $m\ne1$ for $l=2k+1$.

\item {\bf Table~\ref{rez}.} Results of CMB anisotropy spatial
spectrum analysis obtained by different authors basing on COBE
2-years data. For the comparison our result is shown. It is shown
how a quadrupole excluding does influence on the end result.

\end{list}
\pagebreak[4]
\begin{table}
\caption[]{}
\begin{center}
\begin{tabular}{|c|c|c|c|c|c|}
\hline
frequency & $\alpha_5$ & $\Delta\alpha_5$ & $\Delta\alpha_N$ &
			 $\alpha_W$ & $\Delta\alpha_W$ \\

(GHz)	&	     &		       & & & \\
\hline
19.2 -- 31.5 & -2.13 &	 0.36 & 0.38 & -2.12 & 0.69 \\
31.5 -- 53.0 & -2.27 &	 0.28 & 0.36 & -2.19 & 0.76 \\
\hline
\end{tabular}
\end{center}
\label{indx}
\end{table}

\begin{table}
\caption[]{}
\begin{center}
\begin{tabular}{|c|c|c|c|c|}
\hline
frequency & spatial	&$\sigma_{(A+B)/2}$
      & $\sigma_{(A-B)/2}$ & $\sigma_{Sky}$  \\
(GHz)  & filtering &$(\mu K)$ &$(\mu K)$ & $(\mu K)$\\
\hline
\hline
       &	       &       &       &	     \\
       &    yes 	& 641.	& 95.92 &  634.70     \\
\cline{2-5}
31.5   &		       &   &  &  \\
       &	no	      & 254.79& 94.52 & 239.83	\\
\hline
       &	       &       &       &	     \\
       &     no 	& 191.64& 33.04 &  188.80     \\
\cline{2-5}
53.    &		       &   &  &  \\
       &     yes	      &  83.32& 32.20&
76.86 \\
 \hline
\hline
 frequency	      & 		     & & & \\
  "cleaning"          &                      & & & \\

  с $\alpha=-2.15$    &       yes	      &
					     50.78& 45.44& 22.67 \\
\hline
\end{tabular}
\end{center}
\label{tab_gal}
\end{table}

\begin{table}
\caption[]{}
\begin{center}
\begin{tabular}{|c|c|c|c|c|c|c|c|}
\hline
&Quadrupole&$\sigma_{A+B}^2$ & $\sigma_{A-B}^2$ & $\sigma_{Sky}^2$ &
			  $Q_2$
							   & & \\
NN &analysis   &$\mu K^2$   &$ \mu K^2$  &$ \mu K^2$  &$ \mu K$ & $n$ & $M$ \\
\hline
 1 & yes  & $(70.28)^2$ & $(56.76)^2$ & $(41.44)^2 \pm (14.68)^2$ &
$15.22 \pm 2.9$ & $1.84\pm 0.29$  & 446 \\
\hline
 2 & no  & $(68.92)^2$		   &  $(56.47)^2$	     &
 $(39.51)^2 \pm (14.65)^2$  & $15.55\pm 3.8 $ & $1.81\pm 0.37$ & 442 \\
\hline
\hline
 3 & yes  & $(82.91)^2$ & $(74.19)^2$ & $(37.03)^2 \pm (17.48)^2$
 & $18.03 $ & $1.31$ & 649 \\
\hline
 4 & no  &			&			&
			  & $20.3 $ & $1.12$ & 645 \\
\hline
\end{tabular}
\end{center}
\label{tbl_n}
\end{table}

\begin{table}
\caption[]{}
\begin{center}
\begin{tabular}{|c|c|c|}
\hline
Author& $n$ & $Q_{rms-PS}$ \\
\hline
\multicolumn{3}{|c|}{ Quadrupole included} \\
\hline
		  &			   &			  \\
Bennett et al.(1994)& $1.42_{-0.55}^{+0.49}$ & $12.8_{-3.3}^{+5.2} $\\
		  &			   &			  \\
Gorski et al.(1994)& $1.22_{-0.52}^{+0.43}$ & $17.0_{-5.2}^{+7.5}
$\\
		  &			   &			  \\
Wright et al.(1994b)& $1.39_{-0.39}^{+0.34}$ &
\\
		  &			   &			  \\
this work	  & $1.84 \pm 0.29$	   & $15.22 \pm 2.9$	  \\
		  &			   &			  \\
\hline
\multicolumn{3}{|c|}{Quadrupole excluded }\\
\hline
		  &			   &			  \\
Bennett et al.(1994)& $1.11_{-0.55}^{+0.60}$ & $15.8_{-5.2}^{+7.5} $\\
		  &			   &			  \\
Gorski et al.(1994)& $1.02_{-0.59}^{+0.53}$ & $20.0_{-6.5}^{+10.5} $\\
		  &			   &			  \\
Wright et al.(1994a), $l=3-30$& $1.25_{-0.45}^{+0.40}$ &\\
		  &			   &			  \\
Wright et al.(1994b), $l=3-19$& $1.46_{-0.44}^{+0.39}$ &\\
		  &			   &			  \\
this work	  & $1.81 \pm 0.37$	   & $15.55 \pm 3.8$	  \\
		  &			   &			  \\
\hline
\end{tabular}
\end{center}
\label{rez}
\end{table}
\end{document}